\documentclass[journal]{IEEEtran}

\usepackage[pdftex]{graphicx}

\graphicspath{{./pdf/}}

\DeclareGraphicsExtensions{.pdf,.jpeg,.png}

\begin{document}

\title{Luminescence of cadmium fluoride doped with rare-earth ions}

\author{Evgeny Radzhabov and Roman Shendrik ~\IEEEmembership{Member,~IEEE,}
 
\thanks{E.Radzhabov and  R.Shendrik are with Vinogradov Institute of Geochemistry, Russian Academy of Sciences, Favorskii street 1a, P.O.Box 4019, 664033 Irkutsk, Russia and Irkutsk State University, blvd Gagarina 20, Irkutsk, Russia, 664003}

%\thanks{Manuscript received ...}
}

%\markboth{ november 2013}
%\markboth{Journal of \LaTeX\ Class Files,~Vol.~6, No.~1, January~2012}%
%{Shell \MakeLowercase{\textit{et al.}}: Bare Demo of IEEEtran.cls for Journals}

\maketitle

\begin{abstract}

Absorption, excitation and emission spectra of cadmium fluoride crystals doped with rare-earth ions were investigated. In contrast to alkaline-earth fluorides the absorption spectra due to 4f - 5d transitions of Ce$^{3+}$, Pr$^{3+}$ and Tb$^{3+}$ ions are broadened. No 5d-4f emissions were observed. These prove that 5d(e$_g$) levels of rare earth ions lie in conduction band of CdF$_2$ crystal. Emission spectra of Tb$^{3+}$ show the group of 4f-4f $^5$D$_4$-$^7$F$_j$ lines in contrast to other alkaline-earth fluorides where emission due to $^5$D$_3$-$^7$F$_j$ transitions is also observed. The absence of the emission is due to position of $^5$D$_3$ level within condution band.

Among the all measured crystals doped with impurity ions (Pr, Nd, Eu, Ho, Tb, Tm, Yb, Ga, In or Mn), the CdF$_2$ doped with Pr$^{3+}$, Tb$^{3+}$, or Mn$^{2+}$ ions have the highest light outputs under x-ray excitation.

\end{abstract}


\IEEEpeerreviewmaketitle

\section{Introduction}

\IEEEPARstart{C}{admium} fluoride is known as useful material for gamma-ray spectroscopy in 100-300 MeV region with light output near 0.003 relative that of CsI(Tl) \cite{Jones1962}. Later the light output of CaF$_2$ was estimated as 0.13 against that of BGO at room temperature \cite{Derenzo1990}. 
Cadmium fluoride doped with rare-earth and some other trivalent ions can be transformed to semiconductor state by heating in cadmium vapor \cite{Prener1963, Weller1965}. Due to high electron affinity of cadmium the electron from conduction band cannot be trapped directly by trivalent ion or anion vacancy but trapped on donor state of large radius around positive charge \cite{Hayes1974}. 

However, the optical properties of rare-earth ions and particularly the 4f$^n$ - 4f$^{n-1}$5d (hereafte referred as 4f-5d) transitions in this material are poorly known.

In this work the luminescence and ultraviolet absorption spectra of CdF$_2$ undoped and doped with rare-earth ions were investigated. 

\section{Experimental}

The CdF$_2$ crystals doped with trifluorides of Ce, Pr, Nd, Eu, Ho, Tb, Tm, Yb, Ga or difluoride Mn in concentration of 0.01, 0.1 and 1 molar. \% were grown by Stockbarger method in a graphite crucible in vacuum \cite{Radzhabov2012}. 

Absorption spectra were measured using a Perkin-Elmer Lambda 950 spectrophotometer coupled to Janis Research CN-204 cryostat-refrigerator. 
Absorption and excitation spectra in vacuum ultraviolet region were measured using grating monochromator VMR2 (by LOMO) under excitation by a deuterium lamp L7293-50 with MgF$_2$ window (by Hamamatsu) and a solar-blind photomultiplier tube FEU142 as light detector. Emission spectra were measured using grating monochromator MDR2 coupled with a Hamamatsu photo-sensor module H6780-04 operating in counting mode. Spectral resolution for absorption spectra (Figs.1-3) was 0.1-0.2 nm, resolution for emission (Fig.4) was 0.5-1 nm. 

For qualitative evaluation of light output the emission spectra of crystals were measured at room temperatures with 16 nm resolution using 30 kV x-ray tube. 

\section{Results}

The absorption spectra of rare-earth doped cadmium fluoride crystals contain longer-wavelength weak part due to forbidden 4f-4f transitions and shorter-wavelength strong part due to allowed 4f-5d transitions. The 4f-4f absorption spectra are hardly observable in the crystals doped with an impurity concentration less than 1 \%. 

\begin{figure}[!t]
\centering
\includegraphics[width=3.5in]{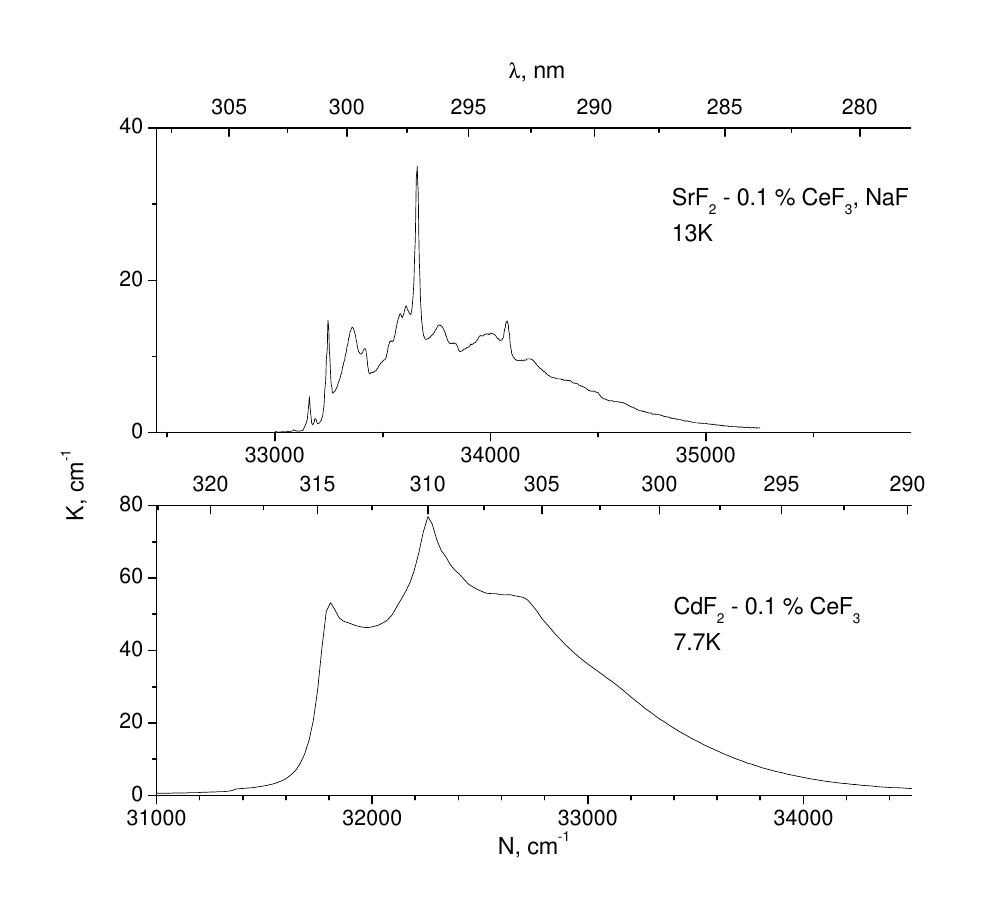}
\caption{Absorption spectra of Ce$^{3+}$ in CdF$_2$ and SrF$_2$ crystals. The spectra were shifted along x-axis to show the similarity.}
\label{Ce-abs}
\end{figure}

\begin{figure}[!t]
\centering
\includegraphics[width=3.5in]{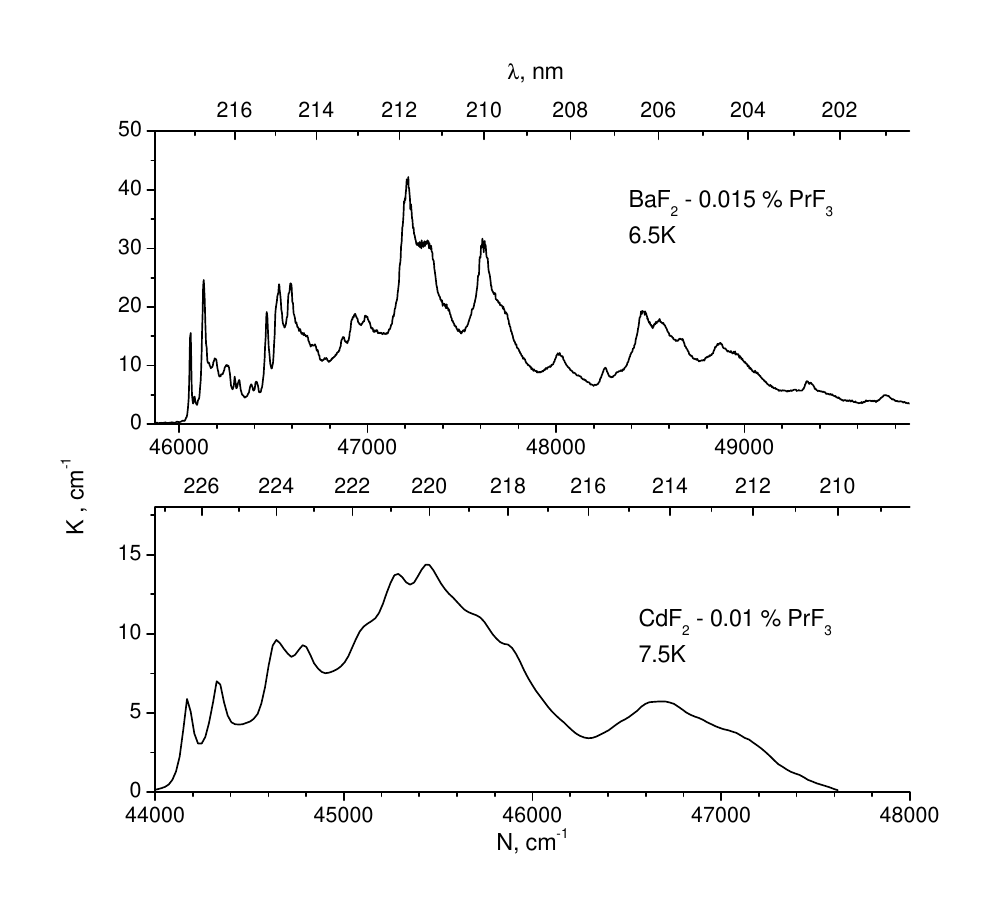}
\caption{Absorption spectra of Pr$^{3+}$ in CdF$_2$ and BaF$_2$ crystals.}
\label{Pr-abs}
\end{figure}

\begin{figure}[!t]
\centering
\includegraphics[width=3.5in]{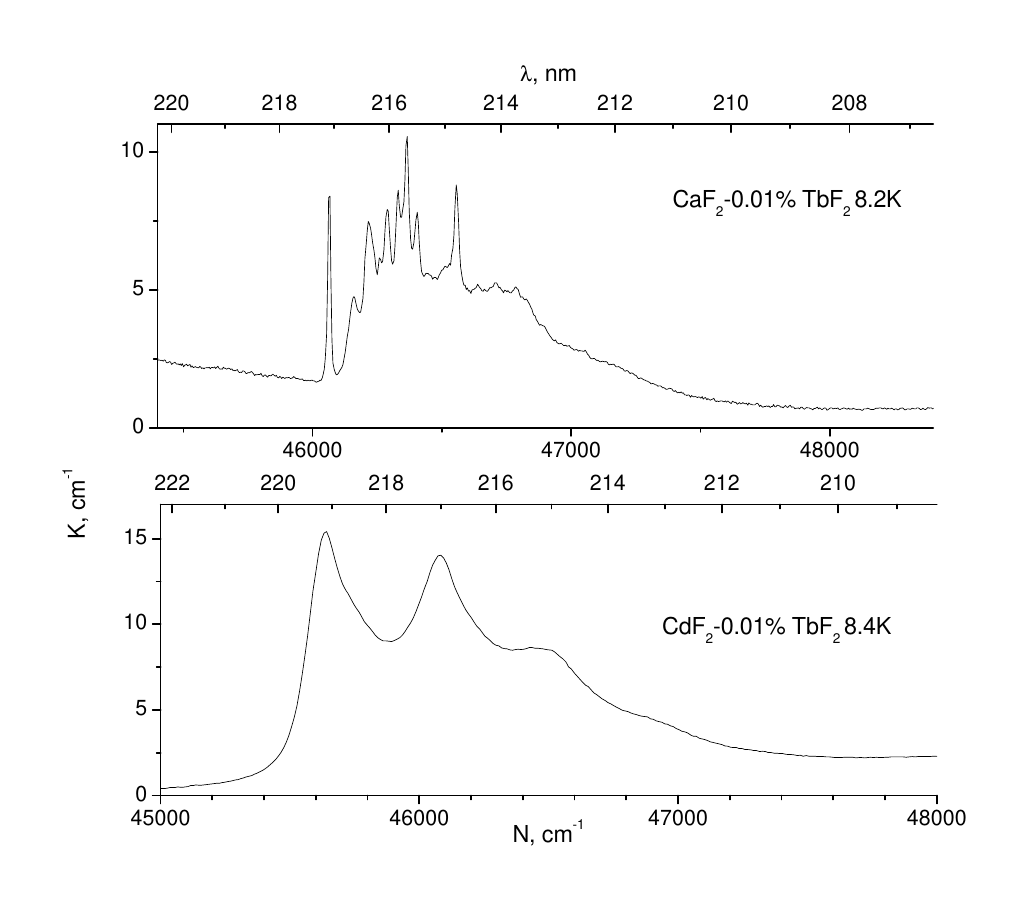}
\caption{Absorption spectra of Tb$^{3+}$ in CdF$_2$ and CaF$_2$ crystals.}
\label{Tb-abs}
\end{figure}

\begin{figure}[!t]
\centering
\includegraphics[width=3.5in]{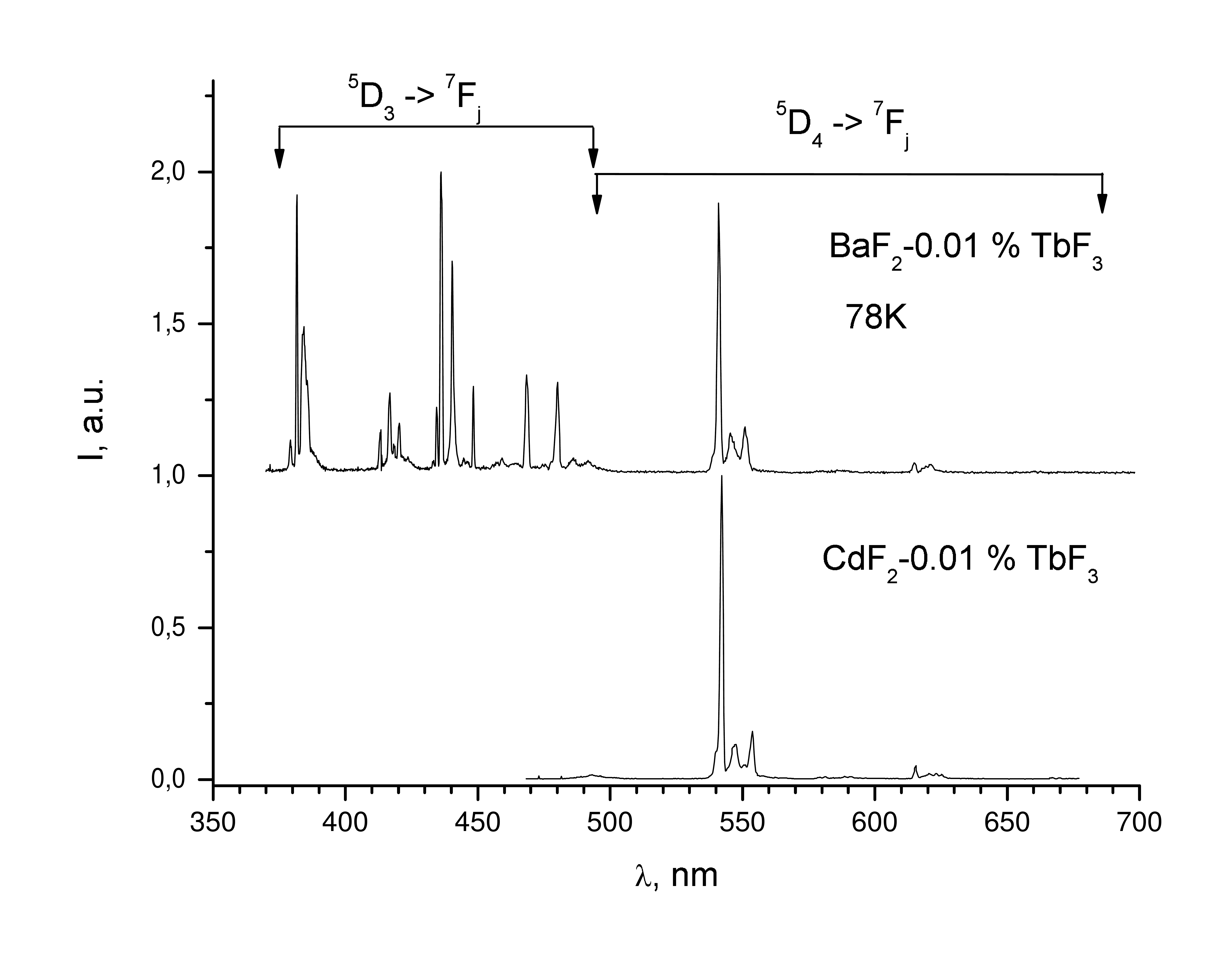}
\caption{Emission spectra of Tb$^{3+}$ in CdF$_2$ and BaF$_2$ crystals measured at 78K. Both samples were excited by 215 nm light.}
\label{Tb-ems}
\end{figure}

The fundamental absorption edge of undoped cadmium fluoride was observed near 200 nm at room temperature, therefore, the 4f-5d absorption bands were found only in Ce$^{3+}$, Pr$^{3+}$ and Tb$^{3+}$ ions doped crystals. For the other rare-earth ions the 4f-5d absorption bands should be observed at shorter wavelengths and they are covered by fundamental absorption of CdF$_2$ host. No 4f$^{n-1}$5d - 4f$^n$ luminescence of Ce$^{3+}$, Pr$^{3+}$, Tb$^{3+}$ ions was observed. 
The absorption lines of  4f - 5d bands of Ce$^{3+}$, Pr$^{3+}$, Tb$^{3+}$ ions in CdF$_2$ were broadened in contrast to the lines in CaF$_2$, SrF$_2$, BaF$_2$ (Fig.\ref{Ce-abs}, \ref{Pr-abs}, \ref{Tb-abs}). 

Positions of the 4f$^n$-4f$^n$ absorption and luminescence lines of trivalent rare earth ions in CdF$_2$ are very similar to those in other alkaline-earth fluorides.  The striking difference was observed for Tb$^{3+}$ luminescence. Two groups of emission lines corresponding to transitions $^5$D$_3$-$^7$F$_j$ (380-480 nm region) and $^5$D$_4$-$^7$F$_j$ (500-650 nm) were observed in alkaline-earth fluorides while only lines due to  $^5$D$_4$-$^7$F$_j$  transitions were observed in CdF$_2$-Tb$^{3+}$ (Fig.\ref{Tb-ems}). 

\begin{figure}[!t]
\centering
\includegraphics[width=3.0in]{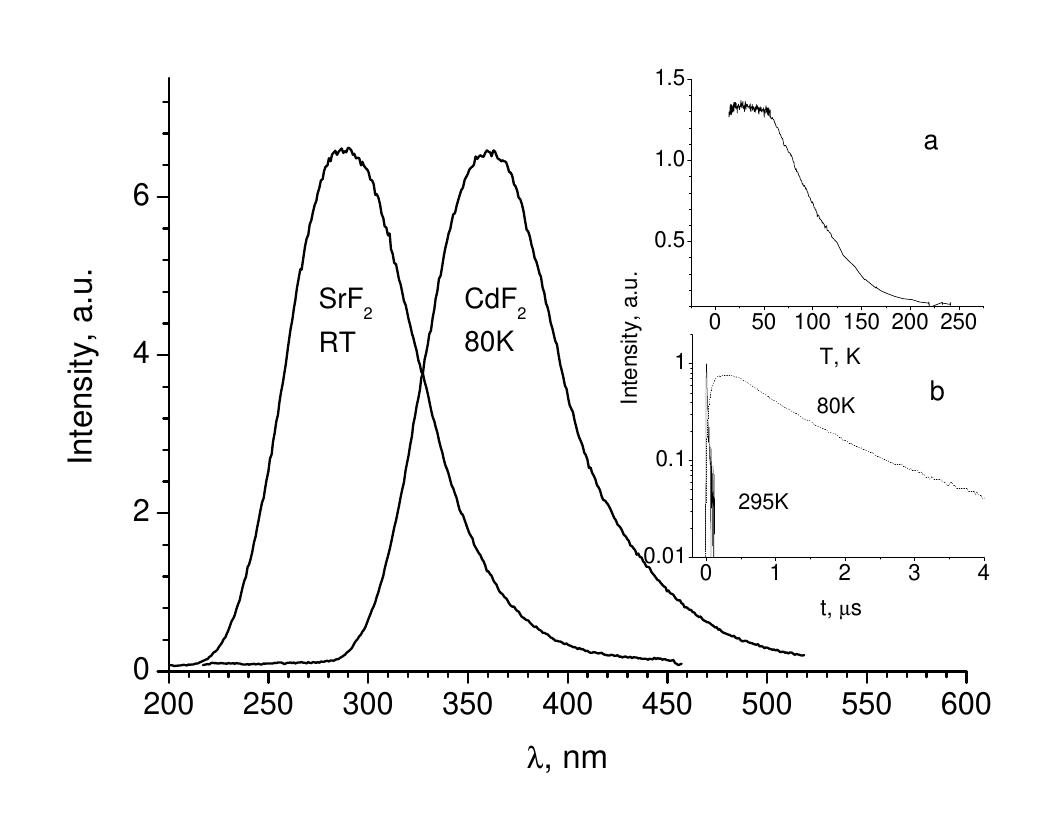}
\caption{Exciton emission spectra of undoped in CdF$_2$ and SrF$_2$ crystals under x-irradiation. The inlet (a) shows the temperature dependence of x-ray induced exciton emission of CdF$_2$ and the inlet (b) show decay of CdF$_2$ exciton emission  at 80K and 295K.}
\label{CdF2-Exciton}
\end{figure}

Exciton emission with wavelength maximum at 370 nm is effectively excited at low temperatures by x-rays (Fig.\ref{CdF2-Exciton}) or by vacuum ultraviolet photons from region 190-140 nm. Intensity of the exciton emission increases with decreasing temperature and reaches the maximum at 60K (see Fig.\ref{CdF2-Exciton}(a)). Excitation spectrum (not shown) has the onset at 192 nm, which correlates with absorption edge at 78K. 
Fast component of decay with tau near 18 ns was observed in all CdF$_2$ crystals at room temperature (see Fig.\ref{CdF2-Exciton}(b)). The emission spectrum of fast component has maximum near 400 nm and obviously belongs to the excitons. Decay time of exciton emission becomes longer with decreasing temperature (see Fig.\ref{CdF2-Exciton}(b)). At least two components of exciton decay with lifetimes 0.8 $\mu$s and 3 $\mu$s were observed at 78K. Comparing the total intensity of x-ray excited emission of CdF$_2$ and SrF$_2$ (Fig.\ref{CdF2-Exciton}) and taking into account the known light output of SrF$_2$ \cite{Shendrik2013}, one could estimate absolute light output of CdF$_2$ at 78K, which is at least 20000 ph/MeV. We consider this value as attainable when the proper luminescent impurity will be doped into the crystal.  

\begin{figure}[!t]
\centering
\includegraphics[width=3.0in]{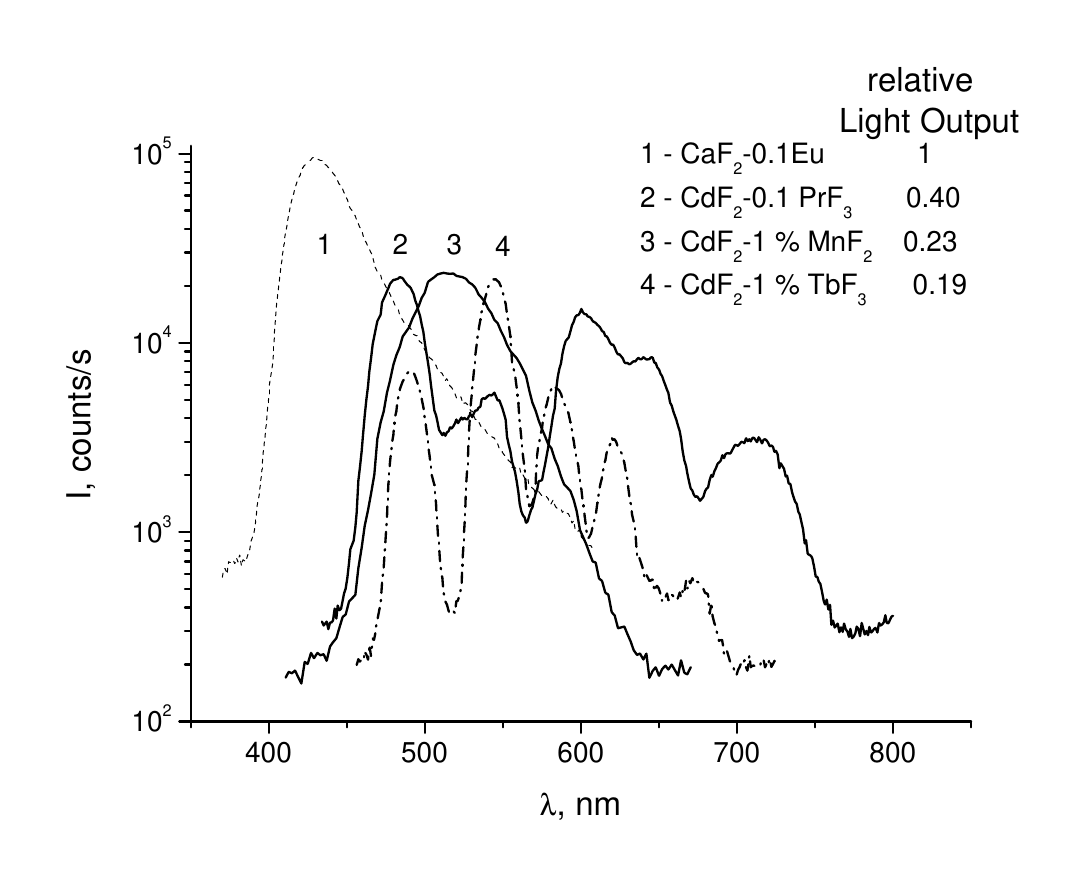}
\caption{X-ray excited emission spectra of CaF$_2$-Eu, CdF$_2$-Pr, CdF$_2$-Mn and CdF$_2$-Tb. Spectra were measured with 16 nm spectral resolution.}
\label{CdF2-output}
\end{figure}

Light output of the doped CdF$_2$ crystals was qualitatively evaluated by comparison of total emission of doped samples at room temperature under x-ray irradiation with that of CaF$_2$-0.1\% EuF$_3$ (Fig.\ref{CdF2-output}).  CaF$_2$-Eu  have known light yield around 24000 ph/MeV \cite{holl1988}. The largest light output was observed for Pr$^{3+}$, Tb$^{3+}$ and Mn$^{2+}$ doped samples. The light output of Pr$^{3+}$ emission (450-770 nm) was estimated as 8000-10000 ph/MeV, which is more than twice less than the light output of excitons. The decay of 4f-4f emission was very slow - 0.3 ms for Pr$^{3+}$ and 9 ms for Tb$^{3+}$ at room temperature. 

\section{Discussion}

On the basis of broadening of absorption lines of divalent europium ion in CdF$_2$ in comparison with sharp lines spectrum in CaF$_2$ it was proved that the excited states of Eu$^{2+}$ are inside conduction band of cadmium fluoride \cite{Godlewski1981}. 
Authors concluded that autoionization process led to broadening of all the optical transitions and smear out any fine structure in the absorption \cite{Godlewski1981}. Another similar model allows to place the 5d(e$_g$) excited levels of all trivalent rare-earth ions at 1.5-2 eV above the bottom of conduction band \cite{Rodnyi2008}. Our measurements show that the absorption spectra of Ce$^{3+}$, Pr$^{3+}$, Tb$^{3+}$ ions in CdF$_2$ are broadened in contrast to the sharp lines in CaF$_2$, SrF$_2$, BaF$_2$ (see Fig.\ref{Ce-abs}, \ref{Pr-abs}, \ref{Tb-abs}). This undoubtedly proves that the lowest 5d(e$_g$) levels of all three ions are located inside the conduction band of CdF$_2$ host. 

From the absence of the luminescence from $^5$D$_3$ level (see Fig.\ref{Tb-ems}) the energy distance from ground  $^7$F$_6$ level of Tb$^{3+}$ to bottom of conduction band should be larger than that of $^7$F$_6$-$^5$D$_4$ transition energy (2.5 eV) but should be less than that of $^7$F$_6$-$^5$D$_3$ (3.3 eV) energy. 
Terbium luminescence in CdF$_2$ was studied in \cite{Bancie1976}. For more than 0.05\% of Tb, the author has only found the fluorescence transitions $^5$D$_4$ - $^7$F$_j$ (with j = 0 to 6). In the presence of Yb or Na the fluorescence transitions starting from the excited level $^5$D$_3$ were observed too \cite{Bancie1976}. Using 4f-5d excitation we clearly show that the $^5$D$_3$ - $^7$F$_j$ emission is absent at lower Tb concentration also (see Fig.\ref{Tb-ems}). 
We also have measured emission spectra of CdF$_2$ doped with 0.1\% and 1\% of TbF$_3$ with excitation into spectral region 337-370  nm ($^7$F$_j$ - $^5$D$_3$ transitions). Besides strong $^5$D$_3$ - $^7$F$_j$ emission we also observed weak $^5$D$_4$ - $^7$F$_j$ emission in CdF$_2$ - 0.1\% TbF$_3$.
The result that perturbation by neighboring Yb or Na leads to appearance of $^5$D$_3$ - $^7$F$_j$ emission \cite{Bancie1976} and our emision spectra under 4f-4f and 4f-5d excitation undoubtedly points on fact, that the $^5$D$_3$ levels of Tb$^{3+}$ locate very close to bottom of conduction band. So one may use the energy of $^7$F$_j$- $^5$D$_3$ transition (3.3 eV) as the energy distance from Tb$^{3+}$ ground state to bottom of conduction band. The experimental value 3.3 eV is slightly larger than 3.0 eV estimated in \cite{Rodnyi2008}. 

Very recently the optical spectra of CdF$_2$-1\% Tb$^{3+}$  were investigated \cite{Boubekri2013}.  The pulled crystals were prepared by use of the Bridgman technique from a vacuum furnace in fluoride atmosphere. The terbium doping is introduced in the form of oxides (Tb$_2$O$_3$) with nominal concentration of 1\% \cite{Boubekri2013}. The emission spectra of sample, illuminated by 350 nm light, exhibit a weak blue emission and a strong green emission in the spectral range 370-460 nm and 478-612 nm, which are assigned to $^5$D$_3$ - $^7$F$_j$ and $^5$D$_4$ - $^7$F$_j$ transitions of Tb$^{3+}$, respectively \cite{Bancie1976}. However all emission bands, published in\cite{Boubekri2013}, were shifted near 10 nm to shorter wavelength side against to those measured in CdF$_2$, CaF$_2$, SrF$_2$, BaF$_2$. It seems the spectra, measured in paper \cite{Boubekri2013}, belong to Tb$^{3+}$ having oxygens in vicinity.

From theoretical calculations \cite{Andriessen1995} the CdF$_2$ crossluminescence (or Auger-free luminescence of core-valence transitions) should be expected around 2-4 eV with a transition time around 100 ns. In \cite{Jones1962} the main emission band of CdF$_2$ extends from 500-580 nm (2.1-2.5 eV) with weaker emission in the blue violet with decay time near 10 ns. However the scintillation efficiency and decay time largerly grew below 150-170K \cite{Jones1962, Rodnyi2001}, while the crossluminescence weakly depends on temperature \cite{Rodnyi2004}. A serious experimental effort was done to search for a fast (∼1 ns) emission in CdF$_2$. However, no ultrafast emission was detected under vacuum ultraviolet or X-ray excitations \cite{Rodnyi2001}. In \cite{Derenzo1990} at least three fast components of 420 nm emission with lifetimes 5, 24 and 78 ns were observed in CdF$_2$ crystals at room temperature. We also observed fast violet emission at room temperature, which becomes slower and much more intense with decreasing temperature. Comparison with emission under vacuum ultraviolet excitation allows us to assign it to exciton emission. 

\section{Conclusion}
The obtained experimental results lead us to the following conclusions:

- first exited 5d(e$_g$) levels of Ce$^{3+}$, Pr$^{3+}$, Tb$^{3+}$ are located inside the conduction band of cadmium fluoride. The energy interval from $^7$F$_0$ ground level of Tb$^{3+}$ to bottom of conduction band equal to 3.3 eV;

- Pr$^{3+}$, Tb$^{3+}$, and Mn$^{2+}$  show the largest light output under x-irradiation due to slow intraconfigurational emission, which is around 6000-10000 ph/MeV,

- fast (nanosecond) x-ray excited emission of CdF$_2$ at room temperature belongs to quenched exciton emission. 

\section*{Acknowledgment}

This work was partially supported by grant 11-02-00717a from Russian Foundation for Basic Research (RFBR). The work was also partially supported by The Ministry of education and science of Russian Federation (N 8382).  The authors are grateful to V. Ivashechkin and V Kozlovskii for the growth of studied crystals.

\bibliographystyle{IEEEtran}
\bibliography{Holmium}

\end{document}